\documentstyle[epsf, rotate]{elsart} 

\begin{document}

\begin{frontmatter}

\title{A Monte Carlo Method for the Numerical Simulation
of Tsallis Statistics}
\author{Rafael Salazar\thanksref{email1}} 
\author{and Ra\'ul Toral\thanksref{email2}}  
\thanks[email1]{e-mail: rafael@imedea.uib.es}
\thanks[email2]{e-mail: raul@imedea.uib.es} 
\address{Instituto Mediterr\'aneo de Estudios Avanzados  
(IMEDEA, UIB--CSIC) and Departament de 
F\'{\i}sica, Universitat de les Illes Balears,  
07071 Palma de Mallorca, Spain. (URL: http://www.imedea.uib.es)}

\begin{abstract}  
We present a new method devised to overcome the intrinsic
difficulties associated to the numerical simulations of the Tsallis statistics.
We use a standard Metropolis Monte Carlo algorithm at a fictitious temperature 
$T'$, combined with a numerical integration method for the calculation of the
entropy in order to evaluate the actual temperature $T$. We illustrate the
method by applying it to the 2d-Ising model using a standard reweighting
technique. 
\end{abstract}


\end{frontmatter}

\section{Introduction}
In 1988 C. Tsallis proposed a thermostatistics formalism based on a
non--extensive entropy definition\cite{tsa88}. In the most recent
formulation\cite{tsa98} of the Tsallis Statistics (TS) in the canonical
ensemble, at fixed temperature $T$, observables ${\cal O}_q$ are obtained as
averages of microscopic functions $O_i$: 
\begin{equation} \label{e3}
{\cal O}_q= \sum_{i=1}^W O_i P_i
\end{equation}
while the entropy is given by
\begin{equation}\label{e1} 
S_q= \frac {(\sum_{i=1}^W P_i^{1/q})^{-q}-1}{1-q} 
\end{equation}
In these expressions, $P_i$ are the escort probabilities\cite{bec93} for 
configuration $i=1,\dots,W$ with energy $\varepsilon_i$:
\begin{equation}\label{e2}
P_i=\frac{A_i}{\sum_{j=1}^W A_j}\equiv\frac {[1-(1-q) \varepsilon_i/T']^{\frac q {1-q}}}  
{\sum_{j=1}^W [1-(1-q) \varepsilon_j/T']^{\frac q {1-q}}} 
\end{equation}
and the additional rule that $A_i=0$ whenever $1-(1-q) \varepsilon_i/T'<0$. 
Here, to simplify notation, one introduces the auxiliary parameter
\begin{equation}\label{e4} 
T'= (1-q)\sum_{i=1}^W \varepsilon_i P_i+
T (\sum_{i=1}^W P_i^{1/q})^{-q}
\end{equation}
The probabilities $P_i$, the entropy $S_q$ and the mean value defining the
observable ${\cal O}_q$ depend, besides the temperature $T$,
on a parameter $q$, which measures the degree 
of non--extensivity of the TS. It is not possible, in general, to solve
the previous equations to 
give explicit expressions for the probabilities $P_i$ as a function
of $q$ and $T$. An exception being the limit $q \to 1$ in which 
one recovers the 
Boltzmann--Gibbs Statistics (BGS): $P_i \propto \exp(- \varepsilon_i/T)$ 
and $S_1= -\sum_{i=1}^W P_i \ln (P_i)$. 
For systems with a small number $W$ of configurations it is possible
to solve equations (\ref{e2},\ref{e4}) iteratively starting from an initial
ansatz for $P_i$, e.g. the Boltzmann-Gibbs expression. However, this method is
not useful for a system with a moderately large number of configurations.

The fact that one can not give explicit expressions for the probabilities $P_i$
has hampered the development of numerical methods to perform the usual
Metropolis Monte Carlo or Molecular Dynamics simulations of the TS for
interacting systems. Very recently, however, methods based on the numerical
calculation of the number of configurations with a given energy have allowed
the direct calculation of the necessary  averages\cite{sal99,lim99}. In this
paper, we introduce a new and more direct method,  based on the standard Metropolis
algorithm combined with a numerical integration, which can be used in many
cases to perform the thermodynamic averages involved in the TS. In the next
section we describe the method in some detail, and in section 3 we show the
results of the application of the method to the Ising model as well as some of
the difficulties encountered.

\section{The Monte Carlo method} 
As mentioned before, the main problem to perform a numerical simulation of a
system  described by the TS at fixed temperature $T$ is that we do  not have at
hand the solution for the probabilities $P_i$, since the nonlinear equations
(\ref{e2},\ref{e4}) have no  explicit solution for $q\ne 1$. For $q=1$ (BGS),
it is $P_i = {\cal Z}^{-1} \exp(-\varepsilon_i/T)$, and one can use a variety
of Monte Carlo techniques for the numerical calculation of the averages, 
Eq.(\ref{e3}). For example: in the Metropolis Monte Carlo algorithm\cite{kalos},
one generates a change in the configuration $i \to j$ and the new 
configuration $j$ is accepted with a probability 
$\min(1,P_j/P_i)=\min[1,\exp(-(\varepsilon_j-\varepsilon_i)/T)]$. Notice that
the partition function ${\cal Z}$ cancels out in the calculation of the
acceptance probabilities. Unfortunately, since for $q \ne 1$ the probabilities
$P_i$ are not known as a function of $T$, there is no trivial generalization of
the Monte Carlo method to perform the averages (\ref{e3}) at fixed temperature
$T$. The method we  propose in this paper works in two steps: (i) we perform
Monte Carlo  simulations at a fixed value of  the auxiliary parameter $T'$;
(ii) we then use Eq.(\ref{e4}) in order to determine the physical temperature
$T$. We describe now in detail both steps.

(i) To perform a Metropolis Monte Carlo simulation of the TS at a fixed 
``fictitious temperature" $T'$, 
one proposes a change in the configuration $i \to j$ and
accepts this change with probability $\min(1,P_j/P_i)$\cite{andre}. Using 
Eq.(\ref{e2}), one notices that the normalizing factor cancels out:
\begin{equation}\label{e5}
P_{accep}=\min \left[1,\left[\frac {T'-(1-q)\varepsilon_j}{T'-(1-q)
\varepsilon_i}\right]^{\frac q {1-q}}\right] 
\end{equation}
 (it is also understood that the acceptance probability is zero if the
configuration $j$ is such that $T'-(1-q)\varepsilon_j<0$). By using this Monte
Carlo algorithm, one generates a sequence of $M$ representative configurations
which are distributed according to the probability Eq.(\ref{e2}). The
statistical averages, Eq.(\ref{e3}), are then approximated by sample averages
${\cal O}_q= \sum_{k=1}^M O_k/M$ and the errors computed in the standard
way\cite{kalos}.

(ii) To perform the $T' \to T$ transformation, we invert Eq.(\ref{e4}) using
Eqs. (\ref{e3},\ref{e1}): 
\begin{equation} \label{e6}
T=\frac{T'-(1-q)U_q(T')}{1+(1-q)S_q(T')} 
\end{equation} 
In this expression, the
energy $U_q(T')$ is obtained in the Monte Carlo  simulation which is performed
at fixed $T'$.  In order to compute the entropy $S_q(T')$ we make use of the 
thermodynamic relation $1/T=\partial S_q /\partial U_q$\cite{pla97b} which,
using Eq.(\ref{e6}), can be integrated between two equilibrium states 
characterized by  values $T'_0$ and $T'$ of the parameter, to yield: 
\begin{equation} \label{e7} \frac 1 {1-q} \ln
\left[\frac{1+(1-q)S_q(T')}{1+(1-q)S_q(T'_0)}\right]  = \int\limits_{T'_0}^{T'}
\frac{d U_q}{T' - (1-q) U_q}  
\end{equation}  
The temperature $T$ is finally
given by: 
\begin{equation} \label{e8}
T=\frac{T'-(1-q)U_q(T')}{1+(1-q)S_q(T'_0)}
\exp\left[(q-1)\int\limits_{T'_0}^{T'} \frac {d U_q}{T' - (1-q) U_q}\right]
\end{equation} 
In summary, one performs a Monte Carlo simulation using the Metropolis
acceptance probability corresponding to states with a fixed value of $T'$,
starting from some initial value $T'_0$ up to the desired valued of $T'$. In
these simulations one computes, using the standard Monte Carlo procedure, the
sample value of the energy $U_q(T')$. Finally, in order to  obtain the physical
temperature $T$ one uses the previous expression Eq.(\ref{e8}) where the
integral is computed numerically. The initial value $T'_0$ must be  some
limiting value in which the entropy $S_q(T'_0)$ is known. This depends on the
problem, although obvious candidates are the very high or very low temperature
configurations.  
One could think that a disadvantage of the method is that the
integration in Eq.(\ref{e8}) requires a large number
of simulations at different fictitious temperatures to be able to perform the
$T' \to T$ transformation accurately. However, we will show that
the use of the reweighting techniques\cite{fer88} reduces drastically the
number of simulations.

\section{The $2$-dimensional Ising Model}
To illustrate the method described in the previous section
we present results for the short range  
Ising Model, defined by the following Hamiltonian:
\begin{equation}\label{e10}
{\cal H} = \sum_{<i,j>} (1-S_i S_j)
\end{equation} 
where each of the $N=L\times L$ spin variables $S_i$ can take the values $\pm 1$, and 
the sum $\sum_{<i,j>}$ runs over all nearest neighbor sites on a 
$2$-dimensional lattice with periodic boundary conditions.

The Metropolis Monte Carlo simulation is performed, as usual, by randomly
choosing one of the spins, $S_i$, and proposing a change in configuration
in which the spin $S_i$ flips its value, $S_i \to -S_i$. This change is accepted
with a probability given by Eq.(\ref{e5}) whereas, if rejected, the spin keeps
its old value. 
\begin{figure}[!ht]
\centerline {\epsfxsize=10cm \epsfbox{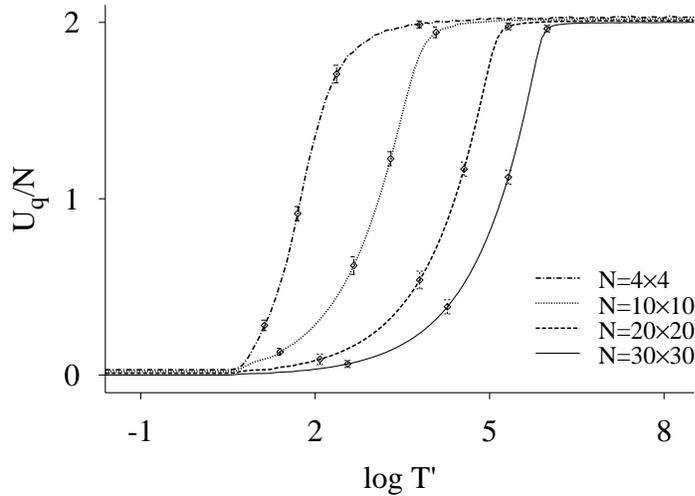}}
\caption{The points show the raw data obtained from
a Metropolis Monte Carlo simulation using Eq.(\ref{e5}). 
The lines connecting the points have been obtained by using a reweighting
extrapolation method, see Eq.(\ref{e11}). We use here $q=0.8$.
}
\label{f1}
\end{figure}

In the Fig.(\ref{f1}) we show the raw Monte Carlo data for $L = 4, 10, 20, 30$,
in the case $q=0.8$. In order to produce this plot, we have used
the Metropolis algorithm at the points marked by symbols in the 
plot. The lines joining the points in the Fig. (\ref{f1}) have been obtained
by a reweighting technique\cite{fer88}, 
which allows the calculation of mean values of observables 
for a $T'$ different from the actual $T'_a$ where the simulation 
was performed. In the case of the TS and the probabilities given by
Eq.(\ref{e1}), the reweighting is based upon the exact relation:
\begin{equation}\label{e11}
{\cal O}_q(T')= \frac {\sum_{\varepsilon} O(\varepsilon)
H(\varepsilon;T'_a) [\frac {1-(1-q) \varepsilon/T'}  
{1-(1-q) \varepsilon/T'_a}]^{\frac q {1-q}}} {\sum_{\varepsilon}
H(\varepsilon;T'_a) [\frac {1-(1-q)\varepsilon/T'}  
{1-(1-q) \varepsilon/T'_a}]^{\frac q {1-q}}} 
\end{equation}
where $H(\varepsilon;T'_a)$ is the histogram of all the energy values generated
in the Monte Carlo run at $T'_a$. Here $O(\varepsilon)$ is the microcanonical
mean value at energy  $\varepsilon$ of the observable $O$. Notice that in 
Eq.(\ref{e11}) the sums run over the energy levels $\varepsilon$ at variance
with Eq.(\ref{e3}) where the sums run over all configurations $i$. 

Once we have obtained the averages as a function of $T'$ as shown in the figure
(\ref{f1}) for the  energy, the value of the temperature $T$ is obtained
from Eq.(\ref{e8}) by performing the indicated integration from the
chosen temperature $T'_0$. In the Ising model, we have used the limit $T'_0=0$
for which the representative configurations are the two ground states, $U_q=0$,
and the entropy is: $S_q(T'_0=0)=\frac {2^{(1-q)}}{1-q}$.

\begin{figure}[!ht]
\centerline {\epsfxsize=10cm \epsfbox{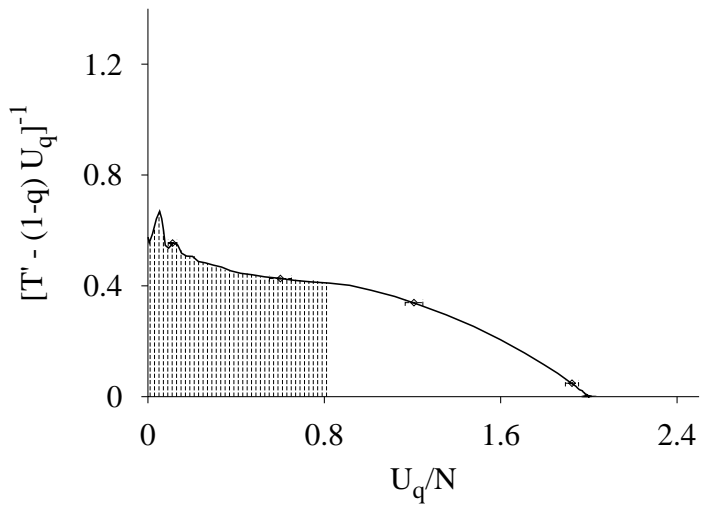}}
\caption{
The shadowed area is the result of the integration needed in Eq.(\ref{e8})
to perform the $T' \to T$ transformation. We
use here $N=10 \times 10$, and $q=0.8$. 
}
\label{f2}
\end{figure}

\begin{figure}[!ht]
\centerline {\epsfxsize=10cm \epsfbox{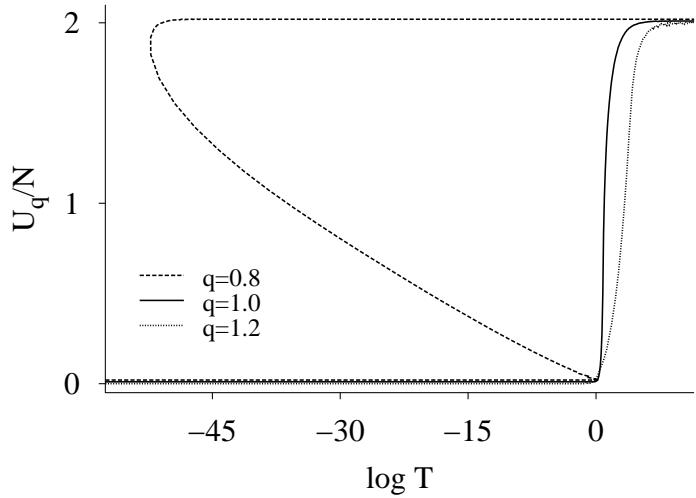}}
\caption{
The internal energy $U_q$ plotted as a function of the actual Temperature $T$
for three values of the parameter $q$. The sizes used here are $N=20 
\times 20$ for $q=0.8,1.0$ and $N=4 \times 4$ for $q=1.2$.
}
\label{f3}
\end{figure}

In Fig.(\ref{f2}) we plot the function that has to be integrated in order to
perform the $T'\to T$ transformation. The main goal of this figure is to show
that the function to integrate is, at least for these values of the parameters
and system sizes, a smooth function. For the actual integration, we have used
Simpson's $3/8$ rule. Finally, in Fig.(\ref{f3}) we plot the resulting
internal energy $U_q$ as a function of the actual temperature $T$, for  several
values of the parameter $q$ and different system sizes. We   note that for
$q<1$ one obtains a  hysteresis--like loop that induces an ambiguity in the
actual value of the  energy. This is a generic feature of the TS, which can be
resolved  by applying  a minimum free energy criterion\cite{lim98} to  choose
the most stable solution. We have observed that for $q>1$ and large system
sizes (this is not the case in Fig. \ref{f3}) the Monte Carlo simulations  in
some temperature ranges near the ferromagnetic transition take a very long time
to equilibrate, thus  preventing us from performing a very accurate
measurement. We believe this is a generic feature of any dynamical updating
scheme one could use to simulate the TS.

Finally, we want to remark that Eq.(\ref{e8}) can be combined with any other
simulation technique which performs a sampling of the configuration space
according to the probability Eq.(\ref{e2}). For instance, one could use the
Molecular Dynamic methods using thermostats at $T'$ \cite{pla97} to study the
dynamic behavior of systems with a large number of degrees of freedom.

\noindent Acknowledgments

We acknowledge financial support
from DGES, grants PB94-1167 and  PB97-0141-C02-01.


\begin{thebibliography}{10}

\bibitem{tsa88}
C.~Tsallis.
\newblock {\em Journal of Statistical Physics}, 52:479, 1988.
For a recent review see: 
C.~Tsallis.
\newblock {\em Brazilian Journal of Physics}, 29:1, 1999.

\bibitem{tsa98}
C.~Tsallis, R.S Mendes, and A.R. Plastino.
\newblock {\em Physica A}, 261:534, 1998.

\bibitem{bec93}
C.~Beck and F.~Schl$\ddot{o}$gl.
\newblock {\sl Thermodynamics of chaotic systems. An introduction}, 
Cambridge Nonlinear Science Series, 1993.

\bibitem{sal99}
R.~Salazar and R.~Toral.
\newblock {\em Physical Review Letters}, (November 1999).

\bibitem{lim99}
A.R. Lima, J.S.~S\'a Martins, and T.J.P. Penna.
\newblock {\em Physica A}, 268:553, 1999.

\bibitem{kalos}
M. Kalos and P. Whitlock.
\newblock {\sl Monte Carlo Methods}, John Wiley and Sons, 1986.

\bibitem{andre}
I.~Andricioaei and J.E. Straub.
\newblock {\em Physical Review E}, 53:R3055, 1996.

\bibitem{pla97b}
A.~Plastino and A.R. Plastino.
\newblock {\em Physics Letters A}, 226:257, 1997.

\bibitem{fer88}
A.M. Ferrenberg and R.H. Swendsen
\newblock {\em Physical Review Letters}, 61:2635, 1988.

\bibitem{lim98}
A.R. Lima and T.J.P. Penna
\newblock {\em Physics Letters A}, 256:221, 1999.

\bibitem{pla97}
A.R. Plastino and C.~Anteneodo.
\newblock {\em Annals of physics}, 255:250, 1997.

\end{thebibliography}
\end{document}